\documentclass[reprint,aip,jap]{revtex4-1}

\usepackage{amssymb}
\usepackage{graphicx}
\usepackage [novolumeabbr] {unitsdef} 

\newunit{\torr}{Torr}	
\newunit{\nanovolt}{nV} 
\newunit{\microvolt}{\mu V} 
\newunit{\Kelvin}{\degree\kelvin} 
\newunit{\gigapascal}{G\pascal} 
\newunit{\ppm}{ppm} 
\newunit{\megapascal}{\mega\pascal} 

\begin{document}
\title {Effects of electron beam induced carbon deposition on the mechanical properties of a micromechanical oscillator}

\author{Stav~Zaitsev}
\email{zzz@tx.technion.ac.il}

\author{Oleg~Shtempluck}
\author{Eyal~Buks}

\affiliation{Electrical Engineering Department, Technion - Israel Institute of Technology, Haifa, Israel 32000}

\begin{abstract}

Electron beam induced deposition of amorphous carbon finds several uses in microlithography, surface micromachining, and the manufacturing of micro- and nanomechanical devices. This process also occurs unintentionally in vacuum chambers of electron microscopes and interferes with normal image acquisition by reducing resolution and causing charging effects. In this work, we show that the resonance frequency of a micromechanical oscillator can be significantly affected by exposing it to a focused electron beam, which induces local carbonization on the surface of the oscillator, resulting in increase in the effective stress along the beam. This \emph{in-situ} carbonization can be utilized for analyzing the amount of residual organic contamination in vacuum chambers. In addition, the method described here allows post-fabrication fine tuning of mechanical resonance frequencies of individual oscillating elements.

\end{abstract}


\maketitle


\section{Introduction}
\label{sec:introduction}

Micromechanical beam structures have long been recognized as effective detectors of chemical and biological materials in microelectromechanical systems (MEMS) \cite{Ekinci_et_al_04a, Fritz_et_al_00, Saeidpourazar&Jalili_09, Yi&Duan_09}.
The adsorption of different chemical substances changes the surface properties of mechanical oscillators, thus affecting their resonance frequencies and other dynamical properties. In particular, surface adsorption affects the internal tension, the effective Young modulus, and the mass distribution along the mechanical beam sensors \cite{Chen_et_al_95, Berger_et_al_97, Cuenot_et_al_04, Lachut&Sader_07, Yi&Duan_09}. A static deformation of a micromechanical sensor due to surface coverage by adsorbates has also been shown \cite{Berger_et_al_97, Fritz_et_al_00, Yu_et_al_09}.
In this paper, we report the effects of electron beam induced deposition (EBID) of amorphous carbon materials on micromechanical beam-string oscillators.

The interaction between an electron beam (e-beam) and the residual organic contamination in the scanning electron microscope (SEM) often results in an unintentional buildup of carbon contamination layers \cite{Reimer_book_85}.
These electrically insulating layers can cause significant degradation of the resolution and the image stability in a SEM. However, the same process can be used constructively in order to deposit masks and create nanoscale structures such as micromechanical clamps \cite{Guise_et_al_04a, Bret_et_al_05, Ding_et_al_05}. The type of materials forming during this procedure is generally known as amorphous hydrogenated carbon \cite{Robertson_02}.
Interestingly enough, these materials exhibit a mixture of diamond-like and graphite-like properties. They often have high mechanical hardness, relatively high Young modulus, and significant internal stresses  \cite{Kelires_93, Pharr_et_al_96, Robertson_02, Ding_et_al_05}.
In this work, we investigate the changes in the dynamical properties of micromechanical oscillators exposed to a focused beam of electrons inside an SEM vacuum chamber.

\section{Experimental setup}
\label{sec:exp_setup}

\begin{figure} [htb]
        \includegraphics [width=3.4in] {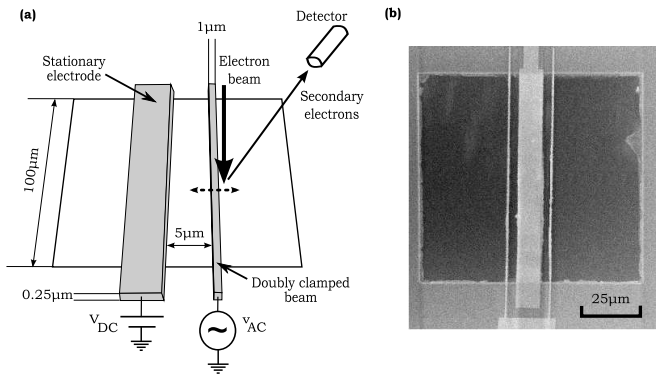}
        \caption{A typical device consists of a suspended doubly clamped narrow beam (length 100-125\micm, width 1\micm, and thickness 0.25\micm) and a wide electrode. The excitation force is applied as voltage between the beam and the electrode. \textbf{(a)} Experimental setup and typical sample's dimensions. The direction of the vibration of the micromechanical beam is denoted by a dotted arrow. In addition, the same dotted arrow shows the direction in which the micromechanical beam-string is continuously scanned by the electron beam during the EBID process. \textbf{(b)} SEM micrograph of a device with one wide electrode and two narrow doubly clamped beams.}
        \label{fig:exp_setup}
\end{figure}

In the experiments, we employ micromechanical oscillators in the form of doubly clamped beams made of $\text{Pd}_{0.15}\text{Au}_{0.85}$ (see Fig.~\ref{fig:exp_setup}).  The dimensions of the beams are: length 100-200\micm, width 1-1.5\micm, and thickness 0.2-0.3\micm; the gap separating the beam and the electrode is 5-8\micm. The fabrication process is described elsewhere \cite{Buks&Roukes_01b}.

The oscillators are excited using capacitive force by applying a combination of DC and AC voltage between the beam-string and the nearby located wide electrode, as shown in Fig.~\ref{fig:exp_setup}. Typical fundamental resonance frequencies of such devices are 150-950\kilohertz, and quality factors are in the range $8000-13000$. Both these parameters depend on the exact manufacturing procedure. In particular, the resonance frequencies are dependent on the relatively high residual tension in the beams \cite{Zaitsev_et_al_11}. Therefore, the micromechanical doubly clamped oscillators used in our experiments are hereafter referred to as beam-strings.

The exposure of micromechanical beam-string oscillators to the electron beam and the measurements of all mechanical properties are done \textit{in-situ} by a SEM imaging system (working pressure $10^{-5}\torr$) \cite{Buks&Roukes_01b}.

During the EBID stage, the SEM electron beam sweeps across the mechanical beam in the transverse direction (see Fig.~\ref{fig:exp_setup}) for a given period of time.

\section{Results}
\label{sec:results}

\begin{figure} [htb]
        \includegraphics [width=3.4in] {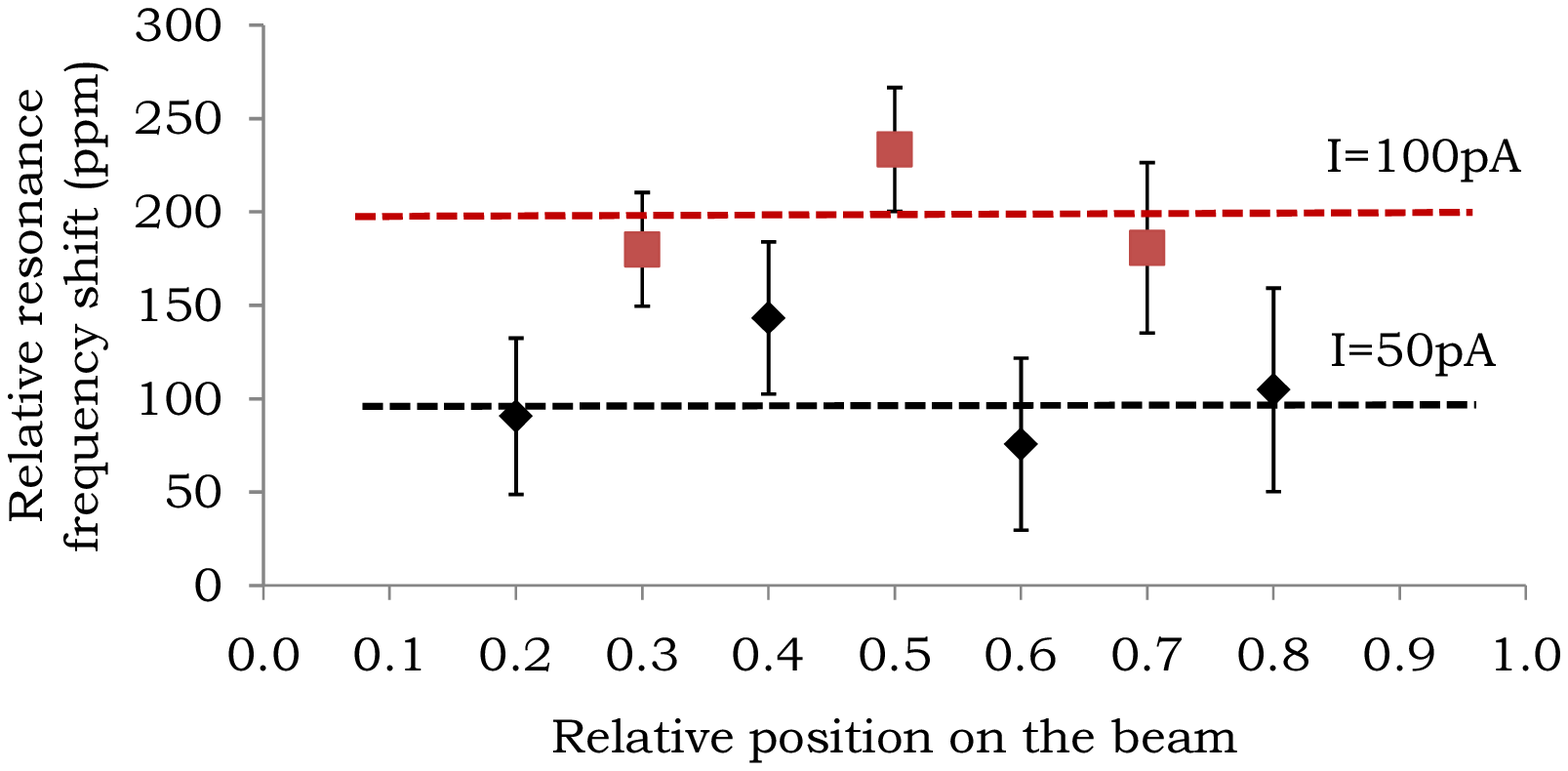}
        \caption{(Color online) Relative resonance frequency shift vs. position of the spot on the beam-string which was exposed to electron beam. Electron accelerating voltage is 10\kilovolt, exposure time is 32\second, electron current is 50\picoampere (black diamonds) and 100\picoampere (brown squares), beam-string length is 125\micrometer, original resonance frequency was 983.2\kilohertz. Dashed lines are drawn as guides to the eye. Note that higher electron beam current results in larger resonance frequency shift. Vertical error bars represent one standard deviation in the measured results.}
        \label{fig:fres_shift_vs_position}
\end{figure}

Micromechanical beam-strings have been exposed to the electron beam at different locations and a shift in resonance frequency has been measured. As can be seen from Fig.~\ref{fig:fres_shift_vs_position}, no significant difference between various exposure locations can be detected, i.e., a shift in the resonance frequency does not depend on the exact location of the exposed spot on the beam-string. This fact suggests that the main reason for resonance frequency shift is an effective change in the internal tension in the beam-string, which could result from a stress in the deposited material. This possibility is further investigated in Sec.~\ref{sec:discussion}.

Interestingly enough, no significant change in the quality factor has been observed in our experiments.

\begin{figure} [htb]
        \includegraphics [width=3.4in] {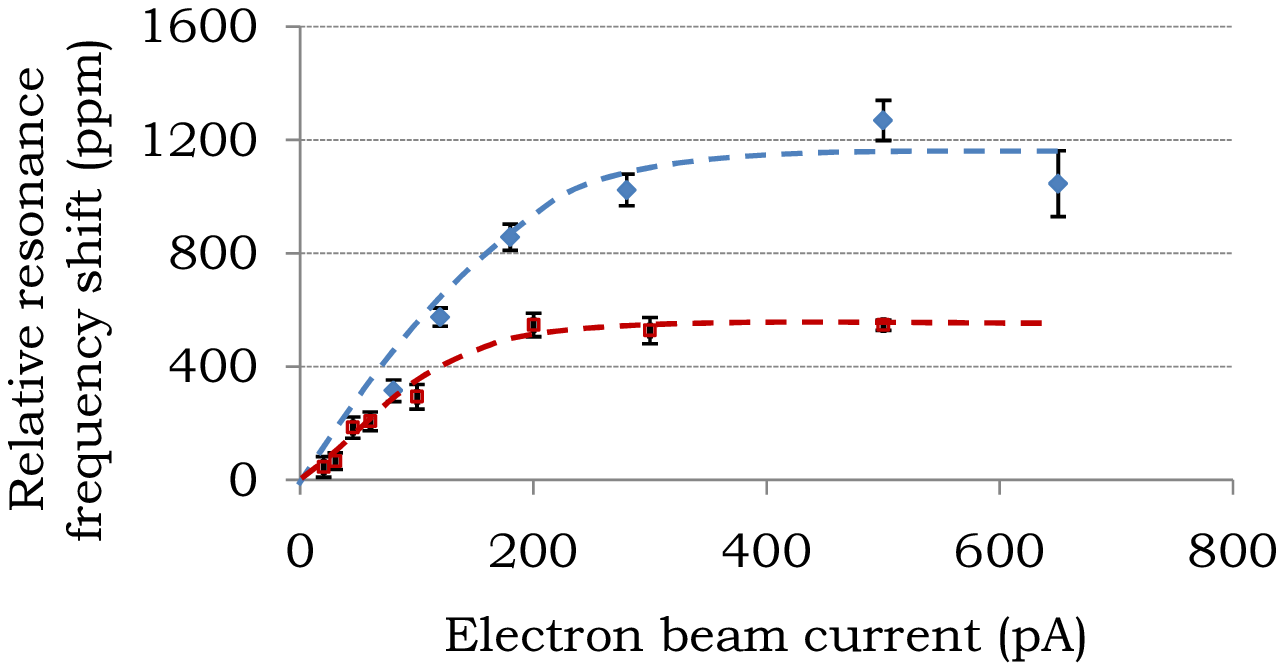}
        \caption{(Color online) Relative resonance frequency shift due to exposure to electron beams with different current magnitudes. Two samples were measured: brown squares correspond to a sample in which the original resonance frequency was 958.3\kilohertz and the exposure time was 20\second; blue diamonds correspond to a sample in which the original resonance frequency was 613.9\kilohertz and the exposure time was 30\second. Electron accelerating voltage is 10\kilovolt,  and the beam-string length in both samples is 125\micrometer. The micromechanical beam-strings were exposed to the electron beam at the relative positions 0.3 and 0.7 along the beam, and the results for these two locations were averaged. Dashed lines are drawn as guides to the eye. Vertical error bars represent one standard deviation in the measured results.}
        \label{fig:fres_shift_vs_current}
\end{figure}

The electron beam current magnitude and the time of exposure are the two most important factors affecting the resonance frequency shift. First, we measure the impact of the e-beam current. The time of exposure is held constant. Typical results are presented in Fig.~\ref{fig:fres_shift_vs_current}. As expected, at low currents ($\lessapprox 200\picoampere$), the resonance frequency shift increases with the current magnitude. However, a saturation occurs at currents above $\approx 300\picoampere$, and further increase in the e-beam current does not result in faster resonance frequency shift.
This saturation behavior can be explained by the finite diffusion rate of precursor organic molecules on the micromechanical beam acting as a limiting factor for EBID \cite{Ding_et_al_05}. This limiting factor may be supplemented by other processes, such as e-beam aided desorption of the EBID products, as explained below.

\begin{figure} [htb]
        \includegraphics [width=3.4in] {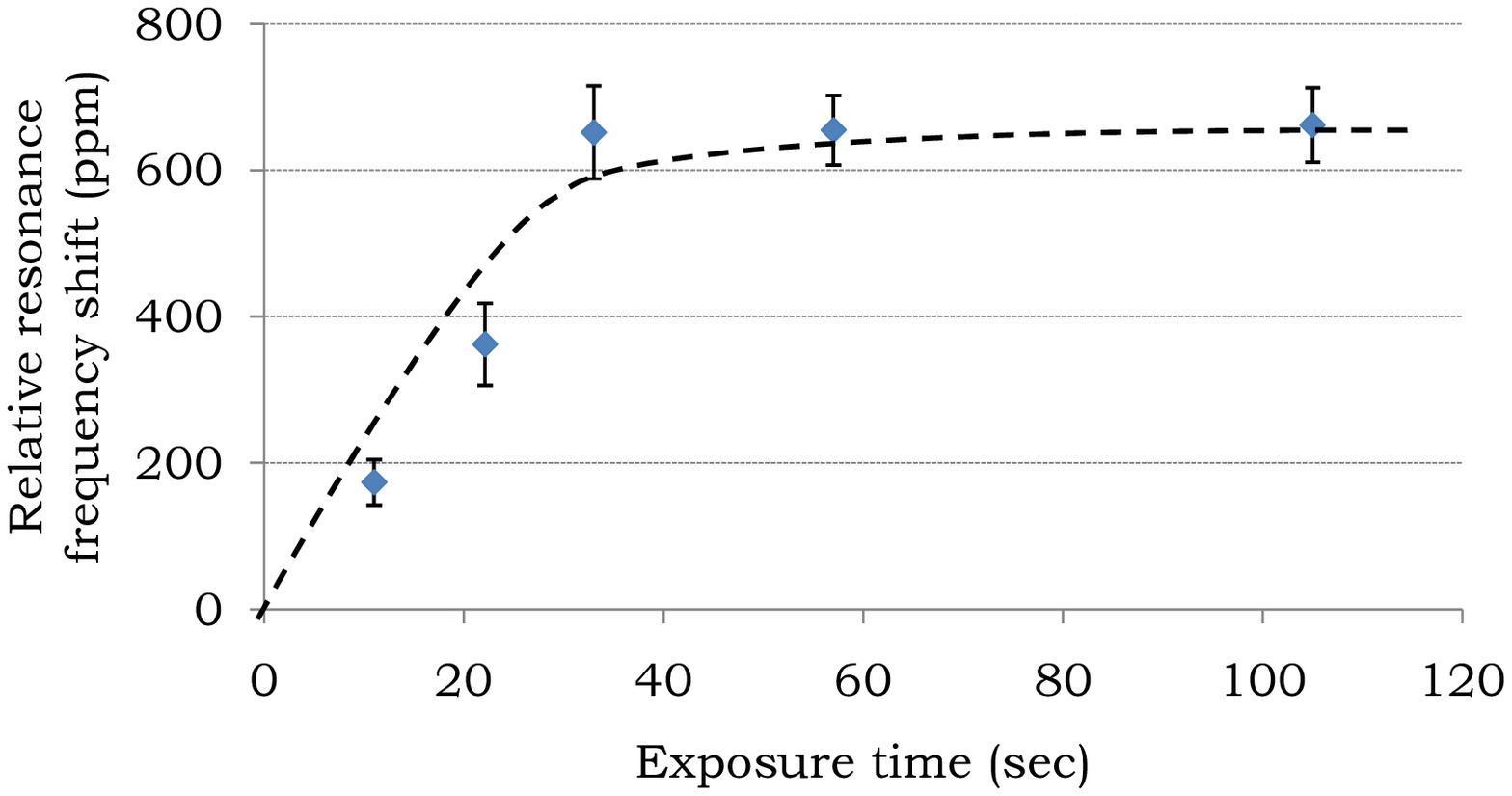}
        \caption{(Color online) Relative resonance frequency shift vs. exposure time. Electron accelerating voltage is 10\kilovolt,  electron current is 300\picoampere, beam-string length is 125\micrometer, original resonance frequency was 953.3\kilohertz. The micromechanical beam-string is exposed to the electron beam at the center. Dashed line is drawn as a guide to the eye. Vertical error bars represent one standard deviation in the measured results.}
        \label{fig:fres_shift_vs_exp_time}
\end{figure}

Next, we measure the impact that the time period of exposure to electron beam has on the resonance frequency shift. Notably, there exist a saturation phenomenon, which is shown in Fig.~\ref{fig:fres_shift_vs_exp_time}.
For exposure periods that are long enough ($\gtrapprox 60\second$), the resonance frequency shift is virtually independent of the actual exposure time. Several explanations are possible. As additional layers of carbon are stacked one above the other, their impact on the effective beam-string center-line tension decreases \cite{Miller&Shenoy_00, Wang&Feng_07}. The EBID method has been shown to result in amorphous carbon materials rich in $sp^2$ bonds \cite{Bret_et_al_05}, and thus having the internal structure resembling graphene. In analogy to graphene layers in graphite, weak interlayer bonds prevent a significant internal stress to exist in the bulk of the deposited amorphous carbon material \cite{Robertson_02}.

Saturation in time is a characteristic of many adsorption-desorption processes~\cite{Berger_et_al_97, Thomas_et_al_91}. We speculate that a similar process can be present in our experiments if the bonds in the products of EBID can be cleaved by either primary or secondary electrons, resulting in highly volatile molecules, which leave the EBID area promptly.

The changes in mechanical properties of the micromechanical resonators in our experiments are partially reversible. After an exposure to atmospheric air, the resonance frequencies of the beam-string are restored to their original values almost completely. Full atmospheric air pressure is not required to achieve this recovery, pressures above $\approx 10^{-2}\torr$ suffice. However, small residual increase in the resonance frequency due to EBID process is permanent. This increase is about 1-5\% of the total resonance frequency change achieved in the experiment prior to loss of vacuum in the SEM chamber. After tens of hours of exposure to electron radiation, interceded by multiple periods of contact with the atmosphere, the permanent change of frequency can reach 20-30\%.

In contrast, after the exposure to e-beam is stopped, the achieved resonance frequency shift does not decrease as long as relatively high vacuum conditions (pressure below $\approx 10^{-4}\torr$) are maintained. No measurable change in the resonance frequency of micromechanical beam oscillators could be detected in our experiments even several hours after the e-beam exposure has ended. Therefore, we conclude that if the saturation phenomena described are indeed a result of an adsorption-desorption equilibrium, both adsorption and desorption processes must be e-beam activated, as no significant desorption can be observed when e-beam is turned off. 

\section{Discussion}
\label{sec:discussion}

Our results contradict the naive view of the local carbonization process solely as a lamped mass addition to the beam-string. Such increase in local mass should result in a decrease in the resonance frequency, which should be position dependent (see, for example, Ref.~\cite{Ilic_et_al_04} and references therein). In contrast, in our experiments, the resonance frequency increases and the increase rate is independent (within experimental error margins) on the position of the spot exposed to the e-beam.

A rough estimation of the added mass due to EBID can be made. The typical density of EBID-produced amorphous hydrogenated carbon \cite{Pharr_et_al_96, Utke_et_al_06} is $1 - 5\gram/\centimeter^3$. For comparison, the density of bulk gold is $19.3\gram/\centimeter^3$. We estimate the volume of the carbon material deposited during one experiment to be less than $10^{-3}\micrometer^3$, which results in total deposited mass of less than $5\times 10^{-15}\gram$. The resultant relative shift in the resonance frequency is estimated to be about $\Delta\omega/\omega_0\approx -10^{-5}=-10\ppm$, which is far smaller in absolute value than the positive changes in resonance frequencies that are experimentally observed.


A simplified model incorporating the phenomena described in this article can be formulated if one considers the change in internal tension in our beam-string mechanical oscillators due to local surface deposition of an excessively stressed thin film. The fundamental resonance frequency of an Euler-Bernoulli beam with significant internal tension can be shown to be given by \cite{Zaitsev_et_al_11} $\omega_0^2=\omega_s^2(1+\pi^2\alpha)$,
where $\omega_s^2=\pi^2N/L^2\rho A$, $\alpha=EI/NL^2$, $N$ is the internal tension, $L$ is the length of the beam-string, $\rho$ is the density of the beam-string, $A$ is the area of the cross section,  $E$ is the Young modulus, and $I$ is the cross section moment of inertia.
Consequently, for small changes in resonance frequency, $\Delta\omega_0/\omega_0=0.5\Delta N/N$, 
where $\Delta\omega_0$ and $\Delta N$ are small changes in resonance frequency and effective beam-string tension respectively. We assume that the Young modulus of the deposited layer is much smaller than the Young modulus of the AuPd alloy from which the micromechanical beam is fabricated.
It follows that the change in the effective internal tension of the beam, $\Delta N$, is approximately equal to the total force applied to the beam by the deposited carbon layer in the longitudinal direction.
Assuming the internal tension $N$ in the micromechanical beam to be of order of 200\megapascal \cite{Zaitsev_et_al_11, Espinosa&Prorok_03},
we estimate the surface tension of the carbon layers grown in our experiments to be of order of $50-100\millinewton\meter^{-1}$. This result is comparable to the published values of surface tension in thin amorphous carbon and other organic layers \cite{Zhao_et_al_04, Chen_et_al_95, Fritz_et_al_00, Yi&Duan_09}.

\section{Summary}
\label{sec:summary}

In summary, we find that the underlying mechanism responsible for the EBID induced frequency shift is the enhancement in the effective tension along the beam.



After an appropriate calibration, this effect can be employed in contamination sensors in SEM vacuum chambers~\cite{Zaitsev_et_al_patent_11}. Although this method does not make a distinction between different contamination materials, it is most responsive to the same materials that cause carbonization in SEM.

The phenomena described above can also be utilized to change the resonance frequencies of micromechanical oscillators, allowing sensitive tuning of these frequencies. Such tuning can be especially useful in arrays of micromechanical beams, 
because each beam can be tuned separately by a local exposure to a focused beam of electrons.

In order to explain the experimental results presented in this paper, further theoretical and experimental work is required. It would be especially interesting to analyze the structure and composition of the deposited materials by chemical and physical means. Due to the highly volatile nature of these materials, analysis \emph{in-situ} SEM vacuum chamber is in order.

\section{Acknowledgment}
The authors are grateful to O. Gottlieb and B. Bar On for helpful discussions. This work is supported by the German Israel Foundation under grant 1-2038.1114.07, the Israel Science Foundation under grant 1380021, the Deborah Foundation, Russell Berrie Nanotechnology Institute, the European STREP QNEMS Project and MAFAT.


\bibliographystyle{unsrt}
\bibliography{EBCD_article_arXiv.bbl}



%

%

\end{document}